# Benefits of Range-separated Hybrid and Double-Hybrid Functionals for a Large and Diverse Dataset of Reaction Energies and Barrier Heights

Golokesh Santra, Rivka Calinsky, and Jan M.L. Martin*

Department of Molecular Chemistry and Materials Science, Weizmann Institute of Science, 7610001 Reḥovot, Israel.

Email: gershom@weizmann.ac.il

**Abstract.** To better understand the thermochemical kinetics and mechanism of a specific chemical reaction, an accurate estimation of barrier heights (forward and reverse) and reaction energies is vital. Due to the large size of reactants and transition state structures involved in real-life mechanistic studies (e.g., enzymatically catalyzed reactions), DFT remains the workhorse for such calculations. In this paper, we have assessed the performance of 91 density functionals for modeling the reaction energies and barrier heights on a large and chemically diverse dataset (BH9) composed of 449 organic chemistry reactions. We have shown that range-separated hybrid functionals perform better than the global hybrids for BH9 barrier heights and reaction energies. Except for the PBE-based range-separated nonempirical double hybrids, range separation of the exchange term helps improve the performance for barrier heights and reaction energies. The sixteen-parameter Berkeley double hybrid, ωB97M(2), performs remarkably well for both properties. However, our minimally empirical range-separated double hybrid functionals offer marginally better accuracy than ωB97M(2) for BH9 barrier heights and reaction energies.

## I. Introduction.

Accurate predictions of kinetic and thermochemical properties are crucial for understanding the mechanisms of different chemical reactions involving main group elements, transition metals, and enzymes.[1,2] By definition, reaction energy (RE) is the energy difference between the product(s) and reactant(s) in their equilibrium state — which has a direct influence on the equilibrium constant of a reaction. On the other hand, barrier heights (BH) are the energy differences between the product(s) or reactant(s) with the transition state (TS). The forward and reverse BHs are the determining components of the reversibility of a reaction.

Traditionally, 1 kcal/mol is considered "chemical accuracy" for bond dissociation energies, heats of reaction, activation barriers, etc. However, a change of approximately 1.4 kcal/mol change in free energy at room temperature results in a change by an order of magnitude for equilibrium constants and reaction rate.[2] Hence, one might choose 1.4 kcal/mol as a "chemical accuracy" criterion for BHs and REs. Highly accurate composite wavefunction *ab initio* methods (for reviews, see ref[3–8] ) can readily achieve this accuracy, but (at least for canonical approaches) their steep computational cost scaling with system size precludes their application to large molecules. As a result, Kohn–Sham density functional theory (KS-DFT[9]) is often seen as a powerful and popular alternative for calculations involving large organic molecules and enzymes.

Depending on the kinds of information employed in the exchange-correlation (XC) functional, Perdew[10] organized DFT methods into what he called a "*Jacob's Ladder*". On each rung of that ladder, dependence on a new type of information is included in the XC functional: the density itself on rung 1 (LDA), the reduced density gradient on rung 2 (GGAs), higher density derivatives or the kinetic energy density on rung 3 (meta-GGAs), occupied orbitals on rung 4, and unoccupied orbitals on rung 5. So-called hybrid and double-hybrid functionals belong to rungs 4 and 5, respectively. In the long-distance limit, the exchange potential of global hybrids deviates from its correct $-1/r_{12}$ ($r_{12}$ being the interelectronic distance) form.[11] Hence, to restore this behavior, the Coulomb operator is partitioned into a short-range (SR) component to be treated by a (meta)GGA, and a long-range (LR) component to be treated by exact exchange, and to 'crossfade' from SR to LR using an error function of $r_{12}$. According to Handy and coworkers[11] the equation has the form:

$$\frac{1}{r_{12}} = \underbrace{\frac{1-[\alpha+\beta \mathrm{erf}(\omega r_{12})]}{r_{12}}}_{\text{SR= Short−Range}} + \underbrace{\frac{\alpha+\beta \mathrm{erf}(\omega r_{12})}{r_{12}}}_{\text{LR= Long−Range}}$$

where the range separation parameter (ω) can either be determined empirically using a training set[11–16] or by minimizing the deviation from the conditions the exact KS functional must obey.[17,18] The parameter α represents the percentage of HF exchange in the short-range limit, and α+β is the corresponding percentage in the long-range limit.

Over the years, several empirical and nonempirical range-separated hybrid (RSH) functionals following the above scheme have been proposed, such as LC-ωPBE,[19] M11,[15] CAM-B3LYP,[11] ωB97X-V,[20] ωB97M-V,[21] and many more. Climbing up one step on the ladder, Ángyán and coworkers[22] and the Head-Gordon group[23] suggested adding a range-separated GLPT2 (second-order Görling−Levy perturbation theory[24]) correction term upon the RSH scheme for accurate long-range correlation energies. However, for their combinatorially optimized, range-separated double hybrid, ωB97M(2),[25] Mardirossian and Head-Gordon instead obtained orbitals from an RSH calculation and then evaluated the GLPT2 correlation in the basis of these orbitals. (ωB97M(2) uses same full semilocal correlation in the orbital generation step as does XYG3[26] and the xDSD functionals considered below. For detailed discussion on DH vs xDH, and on MP2 vs GLPT2 correlation, see refs.[27] and [28], respectively.) Goerigk and coworkers[29–31] and Mester and Kállay[32–35] also proposed and benchmarked several range-separated double hybrid (RSDH) functionals mainly for the electronic excitation energies. For our range-separated dispersion-corrected spin component scaled double hybrids (ωDSD),[36] we used KS orbitals from a standard global hybrid with full semi-local correlation to evaluate the PT2 energies. Adamo *et al.*[37] combined their '*nonempirical*' quadratic integrated double hybrid (QIDH)[38] model with Savin's[39] RSX(range-separated exchange) scheme to propose RSX-QIDH. The range-separation parameter for the RSX-double hybrids were fitted to the exact total ground-state energy of the hydrogen atom. Following the idea of the RSH+MP2[22] method, Kalai and Toulouse[40] proposed a general scheme for RSDHs, where they recommended using range-separation for both exchange and PT2 correlation terms. Recently, Prokopiou *et al.*[41] have developed an optimally tuned RSDH functional by employing the degeneracy-corrected perturbation theory (DCPT2[42]) instead of GLPT2. (We note in passing that the analytical first[43] and second[44] derivatives for DHs are available in the literature, including in continuum solvents.[45] This is of interest not merely for computational spectroscopy, see Refs.[46,47] and references therein, but also will greatly facilitate locating accurate transition state structures.)

Several benchmark studies have shown the excellent performance of range-separated hybrid and double hybrid functionals for calculating the barrier heights and reaction energies involving small and medium-size organic molecules,[36,48–53] transition metals,[36,54–57] and enzymatically catalyzed reactions.[58,59] However, most barrier height and reaction energy datasets available in the literature either focus on only one specific type of reaction, or the reactant molecules are not large enough to represent the systems typically encountered in mechanistic studies.[58,60–68] Hence, assessing quantum chemical methods on kinetic and thermochemical properties based on such databases is prone to bias. Aiming to solve this issue, DiLabio and coworkers[69,70] have recently proposed a large and sufficiently diverse benchmark set, BH9, composed of 449 reaction energies and 898 barrier heights (including forward and reverse). Their dataset contains nine types of reactions: (i) radical rearrangement, (ii) Diels-Alder, (iii) halogen atom transfer, (iv) hydrogen atom transfer, (v) hydride transfer, (vi) B- and Si-containing reactions, (vii) proton transfer, (viii) nucleophilic substitution, and (ix) nucleophilic addition. The reference energies were computed at the DLPNO-CCSD(T)[71–75] [domain-based local pair natural orbital coupled-cluster singles and doubles plus quasiperturbative triples] level of theory at the complete basis set limit.

In Ref.[70], the authors assessed the performance of twenty-five functionals ranging from the first to the fourth rungs of the *Jacob's Ladder.* Other than that, a benchmark study of eighteen double hybrid functionals has recently been published using a reduced version (15 reactions were removed) of BH9 by Brémond *et al.*[76] Interestingly, for both BH and RE, a lower-rung RSH functional, ωB97M-V outperformed the best performing double hybrids recommended in Ref.[76] On top of that, the authors of Ref.[76] considered only a handful of RSDHs, which did not perform well in previous benchmark studies.[36,50,51] Hence, the main objective of the present study is to assess the performance of a variety of range-separated hybrid and double hybrid functionals on BH9 to verify whether range separation is beneficial over the global variants throughout or not. Alongside, we shall explore the effect of systematically increasing the fraction of HF exchange on the performance of global hybrid functionals.

## II. Computational Details.

All electronic structure calculations have been performed with ORCA 5.0.3[77] and QChem 5.4.2[78] running on the Faculty of Chemistry HPC facility. Except for the nucleophilic substitution reactions (i.e., Subset VIII), the Weigend−Ahlrichs family def2-QZVPP[79] basis set has been used throughout. As the reactions of subset VIII contain anions other than hydrides, we have used the minimally augmented diffuse basis set, ma-def2-QZVPP,[80] instead. Appropriate RI[81] basis sets are also employed for the correlation energies. For the ORCA calculations, DEFGRID3 and the RIJCOSX (resolution of the identity in combination with the chain of spheres[82] algorithm) approximation have been used. A pruned integration grid, SG-3,[83] is employed for the QChem calculations. We have considered 91 functionals for the present study ranging from pure GGA (or meta GGA) to range-separated double hybrids. Depending on the exchange and correlation combination used for constructing the functionals, we have divided these 91 functionals into four categories: B97-family, $PBE_x$-$PBE_c$-based, B88-LYP-based, and PBE-P86-based. For the B97-V family functionals, the nonlocal VV10 correction was added in the post-SCF form.

The performance of these functionals is evaluated using the mean absolute deviation (MAD) calculated with respect to the DLPNO-CCSD(T)/CBS reference BHs and REs extracted from ref.[70] See Table S2-S5 in the Supporting Information for the corresponding root mean square deviations (RMSD), mean signed deviations (MSD). The parameters for ωDSD, ωDOD, and their dispersion-free variants can be found in Table S1 of the Supporting Information.

## III. Results and Discussion.

*(a) Barrier heights:*

Table 1 gathers the performance of 91 density functional approximations on the BH9 barrier heights. In general, the range-separated hybrids clearly outperform their global counterparts for B97, $PBE_x$-$PBE_c$, and B88-LYP-based functionals.

Among the B97-family functionals, the BMK-D3BJ global hybrid outperforms the pure meta-GGA functionals B97, B97-D3BJ, and B97M-V for all nine subsets. Except for hydride transfer and nucleophilic substitution reactions, BMK is a better pick than B97-1 for the remaining subsets. With MAD =1.50 and 1.81 kcal/mol, the range-separated hybrids, ωB97X-D and ωB97M-V, perform remarkably well. They even outperform the older range-separated *double* hybrid ωB97X-2 (2.83 kcal/mol), while the top performer among all B97-family functionals is Mardirossian and Head-Gordon's combinatorially optimized range-separated double hybrid ωB97M(2) , with an MAD of just 1.32 kcal/mol.

Let us compare the performance of ωB97M(2) and ωB97M-V for each of the nine reaction types in BH9. For the Diels-Alder, hydride transfer, B & Si-containing reactions, nucleophilic substitution, and nucleophilic additions, ωB97M(2) is the best pick. In contrast, for the remaining four subsets, ωB97M-V wins the race.

If we consider the $PBE_x$-$PBE_c$-based functionals: except for hydride transfer, B & Si-containing reactions, and nucleophilic substitutions, LRC-ωPBEh outperforms its global hybrid counterpart (PBE20) for the remaining barrier heights of BH9. Brémond *et al.*[76] reported the nonempirical double hybrid, PBE0-DH, as their best pick for the reduced BH9 barrier heights. Similarly, among the global double hybrid functionals of this family, PBE0-DH offers the lowest MAD (1.77 kcal/mol) when all of BH9 is considered (898 entries). Adding a D3BJ correction to any of PBE0-DH, PBE-QIDH, or PBE0-2 does more harm than good: detailed inspection of the performance statistics reveals that the dispersion corrections overstabilize the transition state relative to reactant and product, leading to systematic underestimation of barrier heights. It has already been pointed out repeatedly (see Refs.[48,84] and references therein) that if adding dispersion correction adversely affects the performance of a density functional method; it is most likely that the dispersion uncorrected form benefits from error compensation. Hence, although dispersion corrected functional does not offer better accuracy, it paints a "truer picture" of the functional suitability. Nucleophilic substitutions are the only types of reactions where all three dispersion-corrected functionals perform better than the uncorrected counterparts. That being said, for the opposite spin scaled variants of PBE0-DH and PBE-QIDH (i.e., SOS0-PBE-DH and SOS1-PBE-QIDH), dispersion-corrected forms perform better than the respective uncorrected forms. Now, going forward, similar to Refs.[30,31,51] here too the $PBE_x$-$PBE_c$-based range-separated double hybrids are worse performers than the global double hybrids (see Table 1). Closer scrutiny of each subset reveals that range separation of the exchange term is only beneficial for the hydrogen atom and hydride transfer reactions when the QIDH model is considered. On the other hand, hydrogen atom transfer is the only reaction type where RSX-0DH outperforms the PBE-0DH functional.

**Table 1:** Mean absolute deviations (MAD, kcal/mol) of 91 density functionals for full BH9 barrier height set and its nine subsets. The range separation parameters (ω) are also included in a separate column. The nine reaction types of BH9 are radical rearrangement (I), Diels-Alder (II), halogen atom transfer (III), hydrogen atom transfer (IV), hydride transfer (V), Boron and Silicon-containing reactions (VI), proton transfer (VII), nucleophilic substitution (VIII), and nucleophilic addition (IX).

| Family | Functionals | ω | MAD (kcal/mol) | | | | | | | | | |
|---|---|---|---|---|---|---|---|---|---|---|---|---|
| | | | I | II | III | IV | V | VI | VII | VIII | IX | Total |
| B97 based | B97M-V[85] | | 2.41 | 4.77 | 7.50 | 6.18 | 10.18 | 2.75 | 1.38 | 3.59 | 2.82 | 5.18 |
| | B97-D3BJ[86,87] | | 5.54 | 11.17 | 10.94 | 9.78 | 14.36 | 6.17 | 3.49 | 7.75 | 6.31 | 9.61 |
| | B97[86] | | 4.77 | 11.15 | 6.27 | 4.57 | 5.24 | 6.60 | 3.36 | 2.92 | 5.34 | 6.99 |
| | B97-1[88] | | 2.47 | 4.66 | 4.15 | 3.73 | 3.13 | 2.42 | 2.05 | 1.34 | 2.58 | 3.58 |
| | BMK-D3BJ[87,89,90] | | 1.74 | 2.15 | 2.59 | 4.37 | 3.16 | 3.04 | 0.84 | 1.44 | 1.55 | 2.67 |
| | BMK[89] | | 1.46 | 3.12 | 1.95 | 1.94 | 4.84 | 2.31 | 1.56 | 4.09 | 1.44 | 2.59 |
| | ωB97X-D[91] | 0.20 | 0.89 | 1.66 | 1.53 | 3.04 | 1.75 | 1.81 | 0.70 | 2.04 | 0.91 | 1.81 |
| | ωB97X-V[20] | 0.30 | 1.34 | 3.53 | 1.09 | 2.05 | 3.27 | 1.86 | 0.69 | 2.28 | 1.44 | 2.39 |
| | **ωB97M-V[21]** | **0.30** | **0.70** | **1.27** | **1.21** | **1.98** | **3.09** | **1.54** | **0.55** | **1.38** | **0.84** | **1.50** |
| | ωB97X-2-D3BJ[87,92] | 0.30 | 1.96 | 3.97 | 1.80 | 2.54 | 4.48 | 1.36 | 1.36 | 1.48 | 1.67 | 2.83 |
| | ωB97X-2[92] | 0.30 | 1.96 | 3.96 | 1.80 | 2.53 | 4.47 | 1.36 | 1.36 | 1.48 | 1.67 | 2.83 |
| | **ωB97M(2)[25]** | **0.30** | **0.79** | **1.04** | **1.72** | **2.34** | **1.12** | **1.07** | **0.58** | **0.84** | **0.76** | **1.32** |
| PBEx-PBEc based | PBE-D3BJ[87,93,94] | | 5.22 | 10.01 | 12.74 | 12.02 | 16.20 | 6.42 | 6.11 | 7.80 | 6.36 | 10.09 |
| | PBE[93,94] | | 4.52 | 8.00 | 9.80 | 8.50 | 9.87 | 3.80 | 5.48 | 4.35 | 5.17 | 7.41 |
| | PBE20 | | 1.92 | 3.64 | 4.03 | 4.21 | 4.07 | 2.11 | 3.26 | 1.63 | 2.43 | 3.38 |
| | **LRC-ωPBEh[12]** | **0.20** | **1.55** | **2.62** | **2.33** | **2.37** | **4.51** | **2.25** | **2.31** | **4.90** | **1.11** | **2.56** |
| | PBE0-2-D3BJ[95,96] | | 3.68 | 3.81 | 2.19 | 2.57 | 6.83 | 2.30 | 1.67 | 1.02 | 1.85 | 3.30 |
| | PBE0-2[95] | | 3.71 | 3.34 | 2.67 | 2.05 | 5.13 | 2.11 | 1.46 | 1.59 | 1.57 | 2.93 |
| | SOS0-PBE0-2-D3BJ[53,97] | | 4.35 | 2.22 | 4.14 | 1.89 | 1.97 | 1.60 | 0.54 | 1.83 | 1.09 | 2.38 |
| | SOS0-PBE0-2[97] | | 4.64 | 2.60 | 5.63 | 2.82 | 1.87 | 2.90 | 0.84 | 3.83 | 1.11 | 3.03 |
| | PBE0-DH-D3BJ[96,98] | | 2.00 | 2.97 | 1.76 | 3.83 | 4.86 | 2.52 | 1.91 | 0.73 | 2.32 | 2.93 |
| | **PBE0-DH[98]** | | **1.54** | **1.95** | **1.43** | **1.90** | **1.84** | **1.34** | **1.55** | **2.88** | **1.25** | **1.77** |
| | SOS0-PBE0-DH-D3BJ[53,97] | | 1.67 | 2.39 | 1.19 | 2.95 | 3.38 | 2.16 | 1.31 | 1.03 | 1.88 | 2.29 |
| | SOS0-PBE0-DH[97] | | 1.41 | 2.46 | 2.06 | 1.82 | 2.84 | 1.77 | 1.13 | 3.70 | 1.02 | 2.09 |
| | PBE-QIDH-D3BJ[38,99] | | 2.06 | 2.89 | 1.29 | 2.32 | 4.63 | 1.79 | 1.49 | 1.15 | 1.82 | 2.46 |
| | PBE-QIDH[38] | | 2.06 | 2.41 | 1.77 | 1.63 | 2.51 | 1.53 | 1.26 | 2.20 | 1.47 | 2.01 |
| | SOS1-PBE-QIDH-D3BJ[99,100] | | 2.39 | 2.42 | 2.28 | 1.39 | 1.30 | 1.24 | 0.59 | 2.06 | 1.31 | 1.88 |
| | SOS1-PBE-QIDH[100] | | 2.62 | 2.96 | 3.95 | 2.11 | 2.66 | 2.37 | 0.66 | 4.01 | 1.16 | 2.65 |
| | RSX-QIDH-D3BJ[37,51,52] | 0.27 | 3.37 | 5.40 | 3.62 | 1.34 | 1.57 | 2.47 | 1.76 | 3.71 | 3.11 | 3.34 |
| | RSX-QIDH[37,52] | 0.27 | 3.37 | 5.35 | 3.81 | 1.29 | 1.84 | 2.53 | 1.70 | 3.93 | 3.07 | 3.37 |
| | RSX-0DH-D3BJ[51,52] | 0.33 | 3.79 | 7.41 | 5.16 | 1.25 | 6.39 | 3.09 | 1.80 | 6.41 | 3.91 | 4.78 |
| | RSX-0DH[52] | 0.33 | 3.78 | 7.42 | 5.35 | 1.33 | 6.84 | 3.18 | 1.74 | 6.63 | 3.91 | 4.87 |
| | BLYP-D3BJ[87,90,101,102] | | 5.99 | 11.72 | 13.18 | 10.37 | 14.23 | 6.22 | 3.52 | 9.51 | 6.63 | 10.23 |
| | BLYP[101,102] | | 5.20 | 11.60 | 8.23 | 5.15 | 5.37 | 6.55 | 3.36 | 3.68 | 5.40 | 7.52 |
| | B3LYP-D3BJ[87,90,103,104] | | 3.09 | 6.11 | 7.08 | 6.04 | 7.71 | 3.79 | 1.58 | 4.53 | 3.70 | 5.54 |
| | B3LYP[103,104] | | 2.94 | 7.47 | 3.69 | 3.11 | 3.74 | 4.98 | 1.57 | 1.91 | 3.65 | 4.67 |
| | BH&HLYP-D3BJ[87,90,105] | | 1.21 | 3.21 | 2.00 | 1.91 | 4.12 | 2.76 | 2.06 | 2.19 | 1.37 | 2.51 |
| | BH&HLYP[105] | | 1.88 | 6.14 | 5.38 | 5.25 | 11.48 | 5.79 | 3.03 | 5.62 | 2.84 | 5.63 |
| B88-LYP based | **CAM-B3LYP-D3BJ[11,87,90]** | **0.33** | **1.08** | **2.06** | **2.26** | **3.00** | **1.33** | **1.80** | **1.33** | **1.67** | **0.97** | **1.98** |
| | CAM-B3LYP[11] | 0.33 | 1.31 | 3.96 | 2.02 | 2.23 | 6.22 | 3.84 | 1.14 | 3.44 | 1.82 | 3.14 |
| | B2PLYP-D3BJ[87,106] | | 1.65 | 5.73 | 4.96 | 4.48 | 8.08 | 2.05 | 1.28 | 3.32 | 3.00 | 4.56 |
| | B2PLYP[106] | | 1.57 | 5.01 | 2.83 | 2.29 | 3.67 | 2.52 | 1.18 | 1.40 | 2.34 | 3.21 |
| | B2GP-PLYP-D3BJ[87,107] | | 1.28 | 4.42 | 2.43 | 2.88 | 6.20 | 1.03 | 0.86 | 1.81 | 2.07 | 3.19 |
| | B2GP-PLYP[107] | | 1.40 | 3.56 | 1.30 | 1.67 | 3.03 | 1.87 | 0.74 | 1.00 | 1.59 | 2.29 |
| | ωB2PLYP-D3BJ[29,51] | 0.30 | 1.68 | 2.23 | 1.04 | 1.90 | 1.26 | 1.48 | 1.69 | 1.78 | 1.44 | 1.77 |
| | ωB2PLYP[29] | 0.30 | 1.68 | 2.23 | 1.10 | 1.82 | 1.43 | 1.53 | 1.65 | 1.83 | 1.44 | 1.78 |
| | ωB2GP-PLYP-D3BJ[29,51] | 0.27 | 2.12 | 1.94 | 1.40 | 1.67 | 1.05 | 1.43 | 1.46 | 1.42 | 1.35 | 1.67 |
| | **ωB2GP-PLYP[29]** | **0.27** | **2.12** | **1.94** | **1.40** | **1.67** | **1.04** | **1.43** | **1.46** | **1.43** | **1.35** | **1.67** |
| | DSD-PBEP86-D3BJ[108] | | 2.28 | 3.70 | 1.65 | 2.97 | 7.27 | 1.47 | 1.27 | 1.66 | 1.98 | 3.14 |
| | revDSD-PBEP86-D3BJ[109] | | 2.53 | 2.29 | 1.27 | 2.06 | 5.32 | 0.92 | 0.67 | 0.96 | 1.46 | 2.22 |
| | revDOD-PBEP86-D3BJ[109] | | 2.61 | 1.61 | 1.27 | 1.90 | 4.64 | 0.81 | 0.56 | 0.80 | 1.24 | 1.89 |
| | noDispSD-PBEP86[109] | | 2.35 | 4.98 | 1.85 | 3.05 | 7.92 | 1.78 | 1.89 | 1.43 | 2.37 | 3.70 |
| | xDSD$_{75}$-PBEP86-D3BJ[36] | | 2.14 | 2.31 | 1.18 | 2.02 | 5.70 | 0.91 | 0.59 | 0.88 | 1.30 | 2.19 |
| | xDOD$_{75}$-PBEP86-D3BJ[36] | | 2.29 | 1.30 | 1.15 | 1.82 | 4.81 | 0.76 | 0.45 | 0.68 | 1.00 | 1.73 |
| | xnoDispSD$_{75}$-PBEP86[36] | | 1.99 | 4.42 | 1.84 | 2.90 | 7.95 | 1.56 | 1.46 | 1.33 | 2.03 | 3.41 |
| | ωDSD$_{20}$-PBEP86-D3BJ | 0.30 | 0.85 | 0.86 | 1.54 | 2.57 | 1.25 | 1.02 | 1.42 | 1.46 | 0.88 | 1.35 |
| | **ωDOD$_{20}$-PBEP86-D3BJ** | **0.30** | **0.84** | **0.93** | **1.44** | **2.43** | **1.07** | **1.01** | **1.29** | **1.45** | **0.78** | **1.31** |
| | ωnoDispSD$_{20}$-PBEP86 | 0.30 | 1.63 | 4.92 | 4.55 | 5.53 | 6.41 | 2.01 | 4.11 | 2.09 | 2.58 | 4.32 |
| | ωDSD$_{40}$-PBEP86-D3BJ | 0.30 | 1.08 | 0.95 | 1.16 | 2.29 | 2.14 | 0.86 | 1.16 | 1.17 | 0.79 | 1.36 |
| | **ωDOD$_{40}$-PBEP86-D3BJ** | **0.30** | **1.19** | **1.15** | **0.94** | **1.93** | **1.20** | **0.80** | **0.84** | **1.18** | **0.63** | **1.23** |
| | ωnoDispSD$_{40}$-PBEP86 | 0.30 | 1.24 | 3.71 | 2.93 | 4.10 | 5.73 | 1.71 | 2.96 | 1.61 | 1.88 | 3.29 |
| | ωDSD$_{50}$-PBEP86-D3BJ | 0.30 | 1.45 | 1.01 | 0.94 | 2.02 | 2.49 | 0.79 | 0.98 | 1.08 | 0.72 | 1.36 |
| | ωDOD$_{50}$-PBEP86-D3BJ | 0.30 | 1.66 | 1.29 | 0.97 | 1.63 | 1.26 | 0.72 | 0.60 | 1.10 | 0.57 | 1.26 |
| | ωnoDispSD$_{50}$-PBEP86 | 0.30 | 1.37 | 3.13 | 1.96 | 3.35 | 5.35 | 1.65 | 2.38 | 1.39 | 1.51 | 2.80 |
| | ωDSD$_{60}$-PBEP86-D3BJ[36] | 0.22 | 1.46 | 1.32 | 0.99 | 2.14 | 3.60 | 0.79 | 0.83 | 0.85 | 0.94 | 1.59 |
| | **ωDOD$_{60}$-PBEP86-D3BJ[36]** | **0.22** | **1.61** | **0.80** | **0.83** | **1.80** | **2.57** | **0.61** | **0.57** | **0.78** | **0.69** | **1.23** |
| | ωnoDispSD$_{60}$-PBEP86[36] | 0.22 | 1.36 | 4.13 | 2.37 | 3.59 | 6.79 | 1.69 | 2.24 | 1.28 | 1.97 | 3.35 |
| PBE-P86 based | ωDSD$_{60}$-PBEP86-D3BJ | 0.30 | 2.04 | 1.11 | 1.14 | 1.73 | 2.76 | 0.89 | 0.77 | 1.04 | 0.64 | 1.43 |
| | ωDOD$_{60}$-PBEP86-D3BJ | 0.30 | 2.31 | 1.41 | 1.38 | 1.54 | 1.66 | 0.86 | 0.52 | 1.11 | 0.60 | 1.43 |
| | ωnoDispSD$_{60}$-PBEP86 | 0.30 | 1.92 | 2.92 | 1.43 | 2.76 | 5.47 | 1.69 | 1.88 | 1.25 | 1.31 | 2.61 |
| | ωDSD$_{69}$-PBEP86-D3BJ[36] | 0.10 | 1.62 | 1.99 | 1.25 | 2.23 | 5.04 | 0.76 | 0.67 | 0.92 | 1.31 | 2.01 |
| | ωDOD$_{69}$-PBEP86-D3BJ[36] | 0.10 | 1.72 | 1.25 | 1.02 | 1.96 | 4.26 | 0.63 | 0.53 | 0.73 | 1.06 | 1.61 |
| | ωnoDispSD$_{69}$-PBEP86[36] | 0.10 | 1.48 | 4.61 | 2.48 | 3.47 | 7.82 | 1.57 | 1.86 | 1.47 | 2.22 | 3.60 |
| | ωDSD$_{69}$-PBEP86-D3BJ[36] | 0.16 | 1.83 | 1.83 | 1.06 | 2.09 | 4.64 | 0.86 | 0.70 | 0.74 | 1.10 | 1.89 |
| | **ωDOD$_{69}$-PBEP86-D3BJ[36]** | **0.16** | **1.99** | **0.85** | **0.98** | **1.73** | **3.39** | **0.65** | **0.48** | **0.61** | **0.75** | **1.36** |
| | ωnoDispSD$_{69}$-PBEP86[36] | 0.16 | 1.69 | 4.10 | 1.85 | 3.13 | 7.06 | 1.62 | 1.78 | 1.11 | 1.89 | 3.24 |
| | ωDSD$_{69}$-PBEP86-D3BJ | 0.20 | 2.04 | 1.52 | 1.05 | 1.90 | 4.14 | 0.92 | 0.68 | 0.72 | 0.89 | 1.73 |
| | ωDOD$_{69}$-PBEP86-D3BJ | 0.20 | 2.26 | 0.87 | 1.21 | 1.65 | 2.97 | 0.73 | 0.46 | 0.73 | 0.62 | 1.37 |
| | ωnoDispSD$_{69}$-PBEP86 | 0.20 | 1.93 | 3.66 | 1.56 | 2.87 | 6.55 | 1.67 | 1.70 | 1.02 | 1.64 | 2.98 |
| | ωDSD$_{69}$-PBEP86-D3BJ | 0.25 | 2.39 | 1.21 | 1.35 | 1.74 | 3.52 | 1.00 | 0.63 | 0.86 | 0.69 | 1.60 |
| | ωDOD$_{69}$-PBEP86-D3BJ | 0.25 | 2.66 | 1.26 | 1.72 | 1.58 | 2.27 | 0.95 | 0.47 | 1.00 | 0.60 | 1.52 |
| | ωnoDispSD$_{69}$-PBEP86 | 0.25 | 2.27 | 2.89 | 1.36 | 2.40 | 5.57 | 1.66 | 1.49 | 1.05 | 1.23 | 2.54 |
| | ωDSD$_{69}$-PBEP86-D3BJ | 0.30 | 2.76 | 1.31 | 1.73 | 1.72 | 3.32 | 1.16 | 0.64 | 1.09 | 0.64 | 1.70 |
| | ωDOD$_{69}$-PBEP86-D3BJ | 0.30 | 3.10 | 1.63 | 2.22 | 1.57 | 1.83 | 1.19 | 0.48 | 1.30 | 0.69 | 1.72 |
| | ωnoDispSD$_{69}$-PBEP86 | 0.30 | 2.67 | 2.54 | 1.52 | 2.19 | 5.18 | 1.71 | 1.35 | 1.19 | 1.02 | 2.41 |
| | ωDSD$_{72}$-PBEP86-D3BJ[36] | 0.08 | 1.87 | 2.09 | 1.18 | 2.11 | 5.26 | 0.81 | 0.63 | 0.87 | 1.28 | 2.06 |
| | ωDOD$_{72}$-PBEP86-D3BJ[36] | 0.08 | 1.99 | 1.19 | 1.00 | 1.84 | 4.33 | 0.65 | 0.48 | 0.64 | 0.99 | 1.59 |
| | ωnoDispSD$_{72}$-PBEP86[36] | 0.08 | 1.72 | 4.52 | 2.13 | 3.23 | 7.88 | 1.58 | 1.68 | 1.39 | 2.13 | 3.51 |

| | | | | | | | | | | | |
|---|---|---|---|---|---|---|---|---|---|---|---|
| ωDSD$_{72}$-PBEP86-D3BJ[36] | 0.13 | 1.99 | 1.95 | 1.08 | 2.03 | 4.91 | 0.89 | 0.64 | 0.71 | 1.12 | 1.96 |
| ωDOD$_{72}$-PBEP86-D3BJ[36] | 0.13 | 2.17 | 0.93 | 1.08 | 1.74 | 3.80 | 0.71 | 0.45 | 0.57 | 0.80 | 1.46 |
| ωnoDispSD$_{72}$-PBEP86[36] | 0.13 | 1.86 | 4.12 | 1.77 | 2.97 | 7.24 | 1.61 | 1.61 | 1.12 | 1.89 | 3.23 |

Among the B88-LYP-based functionals, global hybrids clearly outperform the pure GGA form. Adding an empirical dispersion correction only helps hybrid functionals with a relatively large percentage of HF exchange (see Table 1). Except for the hydride transfer reactions and nucleophilic substitutions, the range-separated hybrid (i.e., CAM-B3LYP) offers better performance than B3LYP for all other barrier height subsets. However, adding dispersion on top of CAM-B3LYP noticeably improves performance for the aforementioned two subsets of BH9. With MAD=1.98 kcal/mol, CAM-B3LYP-D3BJ even outperforms the higher-rung functionals B2PLYP and B2GP-PLYP. Unlike what we found for the PBE$_x$-PBE$_c$-based functionals, range-separated double hybrids, ωB2PLYP and ωB2GP-PLYP, offer significantly better performance than their global counterparts. Although inclusion of D3BJ degrades the performance for B2PLYP and B2GP-PLYP, it does not affect the statistics for ωB2PLYP and ωB2GP-PLYP. Except for the radical rearrangements, nucleophilic substitutions, halogen atom, and proton transfer reactions, ωB2GP-PLYP offers better performance than the corresponding global DH for the remaining five subsets. In addition to these five reaction types, ωB2PLYP outperforms the B2PLYP functional for halogen atom transfers too.

Turning to the PBE-P86-based functionals, our revised DSD double hybrids clearly outperform the original DSD-PBEP86-D3BJ. Except for the radical rearrangement reactions, only opposite spin scaled, revDOD-PBEP86-D3BJ, performs better than the revDSD variant for the remaining eight subsets. Similar to what we found[109] for the GMTKN55 (general main-group thermochemistry, kinetics, and noncovalent interactions, 55 problem types[48]) benchmark, xDSD$_{75}$-PBEP86-D3BJ and xDOD$_{75}$-PBEP86-D3BJ marginally outperform the corresponding revDSD and revDOD functionals, respectively. However, the revDSD-PBEP86-D3BJ functional outperforms xDSD for hydride transfer reactions. With MAD=1.73 kcal/mol, xDOD$_{75}$-PBEP86-D3BJ is this family's best performing global double hybrid, marginally outperforming the winner of ref.[76] (i.e., PBE0-DH). Like B97 and B88-LYP-family, range-separated PBE-P86-based DHs clearly outperform the global double-hybrid counterparts. With MAD=1.23 kcal/mol, ωDOD$_{40}$-PBEP86-D3BJ (ω=0.3) and ωDOD$_{60}$-PBEP86-D3BJ (ω=0.22) are the two best performers for the full BH9 barrier height dataset and they are even better than ωB97M(2) (1.32 kcal/mol). Except for the proton transfer and nucleophilic substitution reactions, range separation of the exchange part benefits the performance of the seven remaining categories. Now, comparing ωDOD$_{40}$-PBEP86-D3BJ (ω=0.3) and ωDOD$_{60}$-PBEP86-D3BJ (ω=0.22), we found that except for the radical rearrangements, hydride transfer, and nucleophilic addition, the first one performs better than the second functional for all other subsets. For a specific value of ω (e.g., ω=0.3), ωDOD functionals prefer a relatively small fraction of HF exchange at short range when radical rearrangement, Diels-Alder, and hydride transfer reactions are considered. However, for the proton transfer reactions, an ωDOD functional with a large percentage of HF exchange performs better. Performance assessment of ωDOD$_{69}$-PBEP86-D3BJ functional at different "ω" reveals that a relatively large range separation parameter is preferred for the hydrogen atom and hydride transfer reactions. In contrast, for the radical rearrangement, Diels-Alder, B & Si-containing reactions, and nucleophilic substitutions small "ω" performs better. Discarding empirical dispersion correction term does more harm than good for this family of global and range-separated double hybrid functionals.

In response to a reviewer's query, we also evaluated the performance of revPBE and PWPB95 with and without dispersion correction. For BH9 barrier heights, adding D3BJ or D4 correction does more harm than good for both the functionals. With MAD=1.93 kcal/mol, the dispersion-uncorrected global double hybrid PWPB95 offers similar accuracy with revDOD-PBEP86-D3BJ (see Table S6 in the Supporting Information).

*(b)<u>Reaction Energies:</u>*

Table 2 gathers the mean absolute deviations of 91 dispersion-corrected and -uncorrected density functionals for the 449 reaction energies. Similar to what we found for the BH9 barrier heights, the B97, $PBE_x$-$PBE_c$, and B88-LYP-based range-separated hybrids outperform their global hybrid counterparts for BH9 reaction energies.

Among the B97-family functionals, as expected, the global hybrid performs better than the pure mGGA, and range-separated hybrids outperform the global hybrid functionals. Unlike what we found for the BH9 barrier heights, here the ωB97M-V (1.36 kcal/mol) only marginally outperforms the higher-rung functional ωB97X-2 (1.39 kcal/mol). However, with 1.17 kcal/mol mean absolute deviation, ωB97M(2) perform noticeably better than the ωB97M-V. The lion's share of this improvement comes from five subsets: nucleophilic substitutions, B & Si-containing reactions, halogen atom, hydride, and proton transfer reactions.

Turning to the $PBE_x$-$PBE_c$-based functionals: except for the radical rearrangement reactions, LRC-ωPBEh outperforms PBE20 for the remaining eight types of reaction energies. Only opposite spin-scaled $PBE_x$-$PBE_c$-based nonempirical global double hybrid functionals are more efficient than the regular DH counterparts (i.e., PBE0-2, PBE0-DH, and PBE-QIDH) for reaction energies. Similar to our observations for BH9 barrier heights, here exchange range-separation of PBE double hybrids does more harm than good. The three subsets most affected are the radical rearrangement, Diels-Alder, and B & Si-containing reactions. The halogen atom transfer, hydrogen atom transfer, and proton transfer reactions are the only subsets where range-separated DHs offer better accuracy than their global counterparts. SOS1-PBE-QIDH is the best performer (1.81 kcal/mol) among all the functionals tested of this family.

Next, among the B88-LYP-based functionals, range-separated hybrid outperforms the global hybrids functionals when the full BH9 reaction energy set is considered. Adding an empirical dispersion correction helps improving the performance of both global and range-separated hybrids. CAM-B3LYP-D3BJ outperforms the B3LYP-D3BJ for all reaction energies except B & Si-containing reactions. The D3BJ-corrected B88-LYP-based global double hybrids (i.e., B2PLYP-D3BJ and B2GP-PLYP-D3BJ) offer lower MAD values than their uncorrected counterparts. Interestingly enough, range separation of the exchange part only helps for B2PLYP-D3BJ, but for the B2GP-PLYP-D3BJ, it does more harm than good (see Table 2). Three subsets where ωB2PLYP-D3BJ and ωB2GP-PLYP-D3BJ offer better performance than their global DH counterparts are halogen atom, hydrogen atom, and proton transfer reactions. With 1.87 kcal/mol mean absolute deviation, B2GP-PLYP-D3BJ is the best pick among all the B88-LYP family functionals tested in the present study.

**Table 2:** Mean absolute deviations (MAD, kcal/mol) of 91 density functionals for full BH9 reaction energy set and its nine subsets. The range separation parameters (ω) are included in a separate column. The nine reaction types of BH9 are radical rearrangement (I), Diels-Alder (II), halogen atom transfer (III), hydrogen atom transfer (IV), hydride transfer (V), Boron and Silicon-containing reactions (VI), proton transfer (VII), nucleophilic substitution (VIII), and nucleophilic addition (IX).

| Family | Functionals | ω | MAD (kcal/mol) | | | | | | | | | |
|---|---|---|---|---|---|---|---|---|---|---|---|---|
| | | | I | II | III | IV | V | VI | VII | VIII | IX | Total |
| B97 based | B97M-V[85] | | 1.71 | 5.12 | 3.82 | 2.31 | 2.72 | 1.65 | 1.94 | 1.64 | 2.45 | 3.23 |
| | B97-D3BJ[86,87] | | 4.30 | 11.21 | 5.13 | 2.55 | 2.47 | 2.16 | 2.83 | 2.19 | 3.49 | 5.69 |
| | B97[86] | | 7.18 | 19.98 | 6.19 | 2.87 | 2.19 | 10.74 | 2.83 | 2.63 | 9.80 | 9.92 |
| | B97-1[88] | | 2.44 | 8.09 | 3.79 | 2.08 | 2.03 | 3.64 | 1.68 | 1.43 | 2.89 | 4.29 |
| | BMK-D3BJ[87,89,90] | | 2.55 | 3.10 | 1.67 | 1.63 | 2.26 | 4.49 | 1.41 | 1.16 | 2.29 | 2.49 |
| | BMK[89] | | 1.62 | 3.44 | 2.17 | 1.75 | 2.64 | 1.99 | 1.48 | 1.20 | 1.75 | 2.38 |
| | ωB97X-D[91] | 0.20 | 1.24 | 1.76 | 2.31 | 1.61 | 0.93 | 2.50 | 1.20 | 1.19 | 1.28 | 1.65 |
| | ωB97X-V[20] | 0.30 | 2.41 | 3.83 | 1.32 | 1.75 | 0.81 | 3.04 | 0.96 | 1.13 | 2.16 | 2.43 |
| | **ωB97M-V[21]** | **0.30** | **1.04** | **1.46** | **1.50** | **1.21** | **1.34** | **1.98** | **0.73** | **1.11** | **1.23** | **1.36** |
| | ωB97X-2-D3BJ[87,92] | 0.30 | 2.56 | 0.91 | 2.10 | 1.91 | 0.61 | 1.32 | 0.52 | 0.78 | 0.95 | 1.39 |
| | ωB97X-2[92] | 0.30 | 2.56 | 0.91 | 2.10 | 1.91 | 0.61 | 1.32 | 0.52 | 0.78 | 0.95 | 1.39 |
| | **ωB97M(2)[25]** | **0.30** | **1.18** | **1.50** | **0.82** | **1.19** | **0.46** | **1.50** | **0.43** | **0.68** | **1.23** | **1.17** |
| PBEx-PBEc based | PBE-D3BJ[87,93,94] | | 1.89 | 5.35 | 5.63 | 2.91 | 2.81 | 1.39 | 2.87 | 2.74 | 3.61 | 3.73 |
| | PBE[93,94] | | 2.68 | 8.89 | 6.03 | 3.02 | 2.64 | 4.41 | 2.85 | 2.57 | 2.99 | 5.15 |
| | PBE20 | | 2.05 | 4.76 | 3.85 | 2.10 | 1.49 | 3.02 | 1.84 | 1.61 | 1.91 | 3.07 |
| | **LRC-ωPBEh[12]** | **0.20** | **2.89** | **3.21** | **2.91** | **1.88** | **1.14** | **2.41** | **1.27** | **1.19** | **1.52** | **2.42** |
| | PBE0-2-D3BJ[95,96] | | 5.45 | 4.79 | 2.76 | 2.75 | 0.60 | 2.77 | 0.66 | 0.65 | 3.51 | 3.40 |
| | PBE0-2[95] | | 5.35 | 4.10 | 2.77 | 2.73 | 0.62 | 2.11 | 0.65 | 0.52 | 2.95 | 3.09 |
| | SOS0-PBE0-2-D3BJ[53,97] | | 4.31 | 3.58 | 2.65 | 2.45 | 0.55 | 1.84 | 0.52 | 0.54 | 2.07 | 2.67 |
| | SOS0-PBE0-2[97] | | 4.03 | 1.88 | 2.73 | 2.47 | 0.65 | 1.24 | 0.50 | 0.59 | 0.79 | 2.00 |
| | PBE0-DH-D3BJ[96,98] | | 3.91 | 4.98 | 1.79 | 1.42 | 1.16 | 3.29 | 0.95 | 1.22 | 4.42 | 3.11 |
| | PBE0-DH[98] | | 2.96 | 3.07 | 2.18 | 1.45 | 1.28 | 1.24 | 0.96 | 0.95 | 2.22 | 2.17 |
| | SOS0-PBE0-DH-D3BJ[53,97] | | 3.26 | 4.13 | 1.76 | 1.34 | 1.19 | 3.15 | 0.91 | 1.21 | 3.59 | 2.70 |
| | SOS0-PBE0-DH[97] | | 2.50 | 2.80 | 2.15 | 1.39 | 1.36 | 1.60 | 0.91 | 0.97 | 1.53 | 2.02 |
| | PBE-QIDH-D3BJ[38,99] | | 3.78 | 4.71 | 1.41 | 1.77 | 0.84 | 2.33 | 0.67 | 0.71 | 3.54 | 2.87 |
| | PBE-QIDH[38] | | 3.66 | 3.83 | 1.44 | 1.75 | 0.90 | 1.47 | 0.67 | 0.64 | 2.82 | 2.47 |
| | SOS1-PBE-QIDH-D3BJ[99,100] | | 3.04 | 3.91 | 1.29 | 1.59 | 0.86 | 2.07 | 0.57 | 0.75 | 2.54 | 2.41 |
| | **SOS1-PBE-QIDH[100]** | | **2.77** | **2.44** | **1.42** | **1.62** | **1.00** | **1.19** | **0.56** | **0.74** | **1.18** | **1.81** |
| | RSX-QIDH-D3BJ[37,51,52] | 0.27 | 6.11 | 10.04 | 1.33 | 1.58 | 1.31 | 3.50 | 0.61 | 1.21 | 6.19 | 5.03 |
| | RSX-QIDH[37,52] | 0.27 | 6.08 | 9.92 | 1.32 | 1.57 | 1.31 | 3.32 | 0.61 | 1.19 | 6.09 | 4.97 |
| | RSX-0DH-D3BJ[51,52] | 0.33 | 7.55 | 12.18 | 1.57 | 1.14 | 1.88 | 3.96 | 0.80 | 1.65 | 7.06 | 5.95 |
| | RSX-0DH[52] | 0.33 | 7.52 | 12.06 | 1.57 | 1.14 | 1.89 | 3.78 | 0.80 | 1.65 | 6.96 | 5.89 |
| B88-LYP based | BLYP-D3BJ[87,90,101,102] | | 5.55 | 13.30 | 5.21 | 2.74 | 2.34 | 2.50 | 2.67 | 2.61 | 4.15 | 6.59 |
| | BLYP[101,102] | | 7.85 | 20.68 | 6.08 | 2.96 | 2.34 | 10.28 | 2.67 | 2.54 | 9.76 | 10.19 |
| | B3LYP-D3BJ[87,90,103,104] | | 3.33 | 8.18 | 3.11 | 1.83 | 2.11 | 1.61 | 1.69 | 1.80 | 2.61 | 4.14 |
| | B3LYP[103,104] | | 5.15 | 14.27 | 3.85 | 2.06 | 2.34 | 7.66 | 1.68 | 1.59 | 7.11 | 7.10 |
| | BH&HLYP-D3BJ[87,90,105] | | 1.69 | 3.37 | 1.74 | 1.55 | 3.39 | 2.39 | 0.80 | 1.88 | 1.37 | 2.37 |
| | BH&HLYP[105] | | 2.89 | 7.79 | 2.24 | 1.77 | 3.73 | 4.92 | 0.81 | 1.88 | 4.35 | 4.37 |
| | **CAM-B3LYP-D3BJ[11,87,90]** | **0.33** | **1.59** | **3.54** | **2.04** | **1.28** | **1.98** | **1.75** | **0.95** | **1.83** | **1.61** | **2.22** |
| | CAM-B3LYP[11] | 0.33 | 2.36 | 6.72 | 2.41 | 1.42 | 2.24 | 4.50 | 0.93 | 1.47 | 3.56 | 3.70 |
| | B2PLYP-D3BJ[87,106] | | 1.96 | 5.06 | 1.58 | 1.37 | 1.41 | 0.98 | 1.05 | 1.02 | 1.78 | 2.58 |
| | B2PLYP[106] | | 2.69 | 7.99 | 1.95 | 1.44 | 1.54 | 3.83 | 1.04 | 1.07 | 3.76 | 3.97 |
| | B2GP-PLYP-D3BJ[87,107] | | 1.87 | 2.95 | 1.39 | 1.58 | 1.33 | 0.83 | 0.65 | 0.80 | 1.24 | 1.87 |
| | B2GP-PLYP[107] | | 2.06 | 5.01 | 1.52 | 1.63 | 1.41 | 2.48 | 0.65 | 0.78 | 2.24 | 2.75 |
| | **ωB2PLYP-D3BJ[29,51]** | **0.30** | **2.14** | **3.33** | **1.05** | **1.08** | **1.74** | **1.98** | **0.47** | **1.28** | **2.66** | **2.11** |
| | **ωB2PLYP[29]** | **0.30** | **2.13** | **3.27** | **1.05** | **1.08** | **1.74** | **1.87** | **0.47** | **1.26** | **2.60** | **2.08** |
| | ωB2GP-PLYP-D3BJ[29,51] | 0.27 | 2.64 | 3.23 | 1.32 | 1.38 | 1.62 | 1.80 | 0.49 | 1.08 | 2.62 | 2.18 |
| | ωB2GP-PLYP[29] | 0.27 | 2.64 | 3.23 | 1.32 | 1.38 | 1.62 | 1.79 | 0.49 | 1.08 | 2.62 | 2.18 |
| PBE-P86 based | DSD-PBEP86-D3BJ[108] | | 2.94 | 1.09 | 2.07 | 2.10 | 0.62 | 1.51 | 0.68 | 0.83 | 1.59 | 1.58 |
| | revDSD-PBEP86-D3BJ[109] | | 2.59 | 1.05 | 2.14 | 2.04 | 0.60 | 0.86 | 0.68 | 0.66 | 0.87 | 1.43 |
| | revDOD-PBEP86-D3BJ[109] | | 2.48 | 0.93 | 2.10 | 2.00 | 0.61 | 0.83 | 0.66 | 0.65 | 0.90 | 1.37 |
| | noDispSD-PBEP86[109] | | 3.28 | 1.61 | 2.43 | 2.55 | 0.76 | 1.08 | 0.75 | 0.95 | 1.15 | 1.87 |
| | xDSD$_{75}$-PBEP86-D3BJ[36] | | 2.19 | 0.90 | 1.85 | 1.84 | 0.59 | 0.86 | 0.52 | 0.53 | 0.79 | 1.26 |
| | xDOD$_{75}$-PBEP86-D3BJ[36] | | 2.09 | 0.70 | 1.85 | 1.80 | 0.59 | 0.82 | 0.55 | 0.52 | 0.77 | 1.17 |
| | xnoDispSD$_{75}$-PBEP86[36] | | 2.65 | 1.32 | 2.08 | 2.28 | 0.79 | 0.97 | 0.58 | 0.74 | 1.02 | 1.60 |
| | ωDSD$_{20}$-PBEP86-D3BJ | 0.30 | 0.97 | 1.14 | 1.57 | 1.36 | 1.16 | 1.26 | 1.10 | 1.63 | 0.87 | 1.22 |
| | ωDOD$_{20}$-PBEP86-D3BJ | 0.30 | 0.92 | 1.17 | 1.55 | 1.30 | 1.23 | 1.28 | 1.10 | 1.60 | 0.87 | 1.22 |
| | ωnoDispSD$_{20}$-PBEP86 | 0.30 | 1.85 | 1.90 | 2.25 | 2.17 | 1.82 | 1.46 | 1.42 | 1.83 | 1.58 | 1.91 |
| | ωDSD$_{40}$-PBEP86-D3BJ | 0.30 | 1.38 | 1.36 | 1.19 | 1.26 | 0.96 | 1.26 | 0.85 | 1.23 | 0.84 | 1.23 |
| | ωDOD$_{40}$-PBEP86-D3BJ | 0.30 | 1.31 | 1.53 | 1.15 | 1.18 | 1.04 | 1.24 | 0.89 | 1.19 | 0.82 | 1.26 |
| | ωnoDispSD$_{40}$-PBEP86 | 0.30 | 2.12 | 1.69 | 1.64 | 1.89 | 1.33 | 1.24 | 1.02 | 1.31 | 1.35 | 1.65 |
| | ωDSD$_{50}$-PBEP86-D3BJ | 0.30 | 1.79 | 1.53 | 1.16 | 1.31 | 0.86 | 1.29 | 0.70 | 0.99 | 0.89 | 1.32 |
| | ωDOD$_{50}$-PBEP86-D3BJ | 0.30 | 1.66 | 1.71 | 1.12 | 1.21 | 0.93 | 1.21 | 0.83 | 0.99 | 0.84 | 1.34 |
| | ωnoDispSD$_{50}$-PBEP86 | 0.30 | 2.46 | 1.66 | 1.52 | 1.92 | 1.11 | 1.24 | 0.81 | 1.05 | 1.28 | 1.64 |
| | ωDSD$_{60}$-PBEP86-D3BJ[36] | 0.22 | 1.67 | 1.06 | 1.23 | 1.33 | 0.70 | 1.07 | 0.68 | 0.79 | 0.77 | 1.13 |
| | **ωDOD$_{60}$-PBEP86-D3BJ[36]** | **0.22** | **1.56** | **1.12** | **1.19** | **1.28** | **0.74** | **1.00** | **0.71** | **0.79** | **0.76** | **1.12** |
| | ωnoDispSD$_{60}$-PBEP86[36] | 0.22 | 2.30 | 1.41 | 1.55 | 1.98 | 1.02 | 1.12 | 0.80 | 0.99 | 1.18 | 1.53 |
| | ωDSD$_{60}$-PBEP86-D3BJ | 0.30 | 2.35 | 1.82 | 1.54 | 1.55 | 0.74 | 1.35 | 0.56 | 0.79 | 0.98 | 1.55 |
| | ωDOD$_{60}$-PBEP86-D3BJ | 0.30 | 2.23 | 2.08 | 1.56 | 1.47 | 0.87 | 1.30 | 0.75 | 0.81 | 0.94 | 1.61 |
| | ωnoDispSD$_{60}$-PBEP86 | 0.30 | 2.92 | 1.69 | 1.87 | 2.17 | 1.01 | 1.30 | 0.66 | 0.87 | 1.20 | 1.76 |
| | ωDSD$_{69}$-PBEP86-D3BJ[36] | 0.10 | 1.68 | 0.93 | 1.40 | 1.51 | 0.59 | 0.82 | 0.65 | 0.58 | 0.78 | 1.11 |
| | **ωDOD$_{69}$-PBEP86-D3BJ[36]** | **0.10** | **1.59** | **0.78** | **1.38** | **1.48** | **0.60** | **0.80** | **0.63** | **0.56** | **0.78** | **1.04** |
| | ωnoDispSD$_{69}$-PBEP86[36] | 0.10 | 2.24 | 1.43 | 1.65 | 2.03 | 0.83 | 0.99 | 0.74 | 0.84 | 1.11 | 1.51 |
| | ωDSD$_{69}$-PBEP86-D3BJ[36] | 0.16 | 1.99 | 0.87 | 1.57 | 1.62 | 0.60 | 0.99 | 0.59 | 0.58 | 0.74 | 1.17 |
| | ωDOD$_{69}$-PBEP86-D3BJ[36] | 0.16 | 1.87 | 0.96 | 1.51 | 1.52 | 0.62 | 0.94 | 0.63 | 0.58 | 0.77 | 1.16 |
| | ωnoDispSD$_{69}$-PBEP86[36] | 0.16 | 2.52 | 1.28 | 1.81 | 2.14 | 0.85 | 1.06 | 0.66 | 0.80 | 1.10 | 1.54 |
| | ωDSD$_{69}$-PBEP86-D3BJ | 0.20 | 2.25 | 1.13 | 1.69 | 1.69 | 0.62 | 1.12 | 0.52 | 0.60 | 0.79 | 1.32 |
| | ωDOD$_{69}$-PBEP86-D3BJ | 0.20 | 2.13 | 1.29 | 1.67 | 1.60 | 0.68 | 1.07 | 0.64 | 0.61 | 0.79 | 1.34 |
| | ωnoDispSD$_{69}$-PBEP86 | 0.20 | 2.77 | 1.32 | 1.97 | 2.24 | 0.88 | 1.13 | 0.62 | 0.79 | 1.10 | 1.62 |
| | ωDSD$_{69}$-PBEP86-D3BJ | 0.25 | 2.62 | 1.63 | 1.86 | 1.80 | 0.65 | 1.30 | 0.49 | 0.61 | 0.94 | 1.58 |
| | ωDOD$_{69}$-PBEP86-D3BJ | 0.25 | 2.53 | 1.99 | 1.85 | 1.71 | 0.73 | 1.27 | 0.69 | 0.62 | 0.99 | 1.67 |
| | ωnoDispSD$_{69}$-PBEP86 | 0.25 | 3.10 | 1.50 | 2.13 | 2.31 | 0.85 | 1.25 | 0.56 | 0.71 | 1.11 | 1.74 |
| | ωDSD$_{69}$-PBEP86-D3BJ | 0.30 | 3.06 | 2.08 | 2.10 | 2.01 | 0.69 | 1.46 | 0.52 | 0.61 | 1.08 | 1.85 |
| | ωDOD$_{69}$-PBEP86-D3BJ | 0.30 | 2.92 | 2.49 | 2.08 | 1.88 | 0.81 | 1.41 | 0.75 | 0.62 | 1.10 | 1.95 |
| | ωnoDispSD$_{69}$-PBEP86 | 0.30 | 3.47 | 1.83 | 2.36 | 2.46 | 0.89 | 1.39 | 0.53 | 0.68 | 1.20 | 1.96 |
| | ωDSD$_{72}$-PBEP86-D3BJ[36] | 0.08 | 1.93 | 0.88 | 1.61 | 1.67 | 0.58 | 0.85 | 0.59 | 0.56 | 0.78 | 1.17 |
| | **ωDOD$_{72}$-PBEP86-D3BJ[36]** | **0.08** | **1.83** | **0.71** | **1.58** | **1.61** | **0.58** | **0.82** | **0.59** | **0.53** | **0.78** | **1.09** |
| | ωnoDispSD$_{72}$-PBEP86[36] | 0.08 | 2.45 | 1.35 | 1.86 | 2.17 | 0.82 | 0.99 | 0.66 | 0.80 | 1.08 | 1.55 |
| | ωDSD$_{72}$-PBEP86-D3BJ[36] | 0.13 | 2.13 | 0.84 | 1.70 | 1.73 | 0.58 | 0.97 | 0.54 | 0.55 | 0.75 | 1.20 |
| | ωDOD$_{72}$-PBEP86-D3BJ[36] | 0.13 | 2.02 | 0.88 | 1.67 | 1.65 | 0.59 | 0.91 | 0.60 | 0.52 | 0.76 | 1.18 |
| | ωnoDispSD$_{72}$-PBEP86[36] | 0.13 | 2.61 | 1.24 | 1.95 | 2.23 | 0.81 | 1.04 | 0.62 | 0.78 | 1.06 | 1.56 |

Now, if we consider the PBE-P86-based global and range-separated DHs, revDSD-PBEP86-D3BJ clearly outperforms the original DSD-PBEP86-D3BJ functional. Similar to BH9 barrier heights, here, revDOD-PBEP86-D3BJ and $xDOD_{75}$-PBEP86-D3BJ offer better accuracy than their DSD counterparts. A significant share of this performance improvement comes from the radical rearrangement and Diels-Alder reactions. If all 449 reaction energies are considered, $xDOD_{75}$-PBEP86-D3BJ performs similarly to the range-separated Berkeley double hybrid ωB97M(2) (see Table 2). Comparing these two functionals for nine subsets, we found that except for Diels-Alder, B & Si-containing reactions, nucleophilic substitution, and addition reactions, ωB97M(2) outperforms global double hybrid $xDOD_{75}$-PBEP86-D3BJ for all other subsets. A number of our range-separated ωDSD functionals offer better accuracy than the best global double hybrid, $xDOD_{75}$-PBEP86-D3BJ of this family. Hence, range separation of the exchange part of our DSD-family double hybrids clearly benefits for the BH9 reaction energies. Unlike for BH9 barrier heights, the overall performance of the ωDSD and ωDOD-PBEP86-D3BJ functionals are comparable for reactions energies. With 1.04 kcal/mol mean absolute deviation, the $\omega DOD_{69}$-PBEP86-D3BJ (ω=0.10) is the best pick among the PBE-P86 family as well as all the functionals tested in the present study for BH9 reactions energies. Comparing the performance of best global and range-separated DHs of this family, we found that radical rearrangements, halogen, and hydrogen atom transfer reactions enjoy the lion's share of the benefit from range separation.

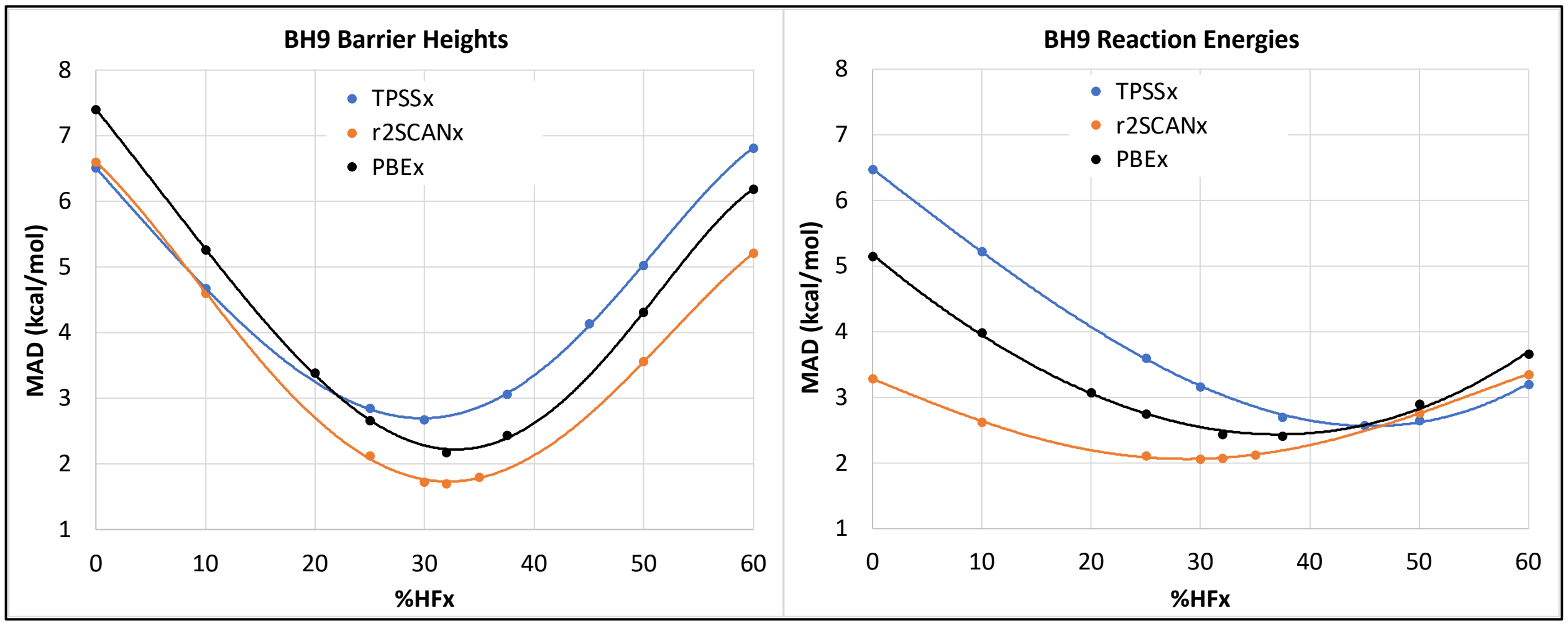

**Figure 1:** Dependence of mean absolute deviations (MAD, in kcal/mol) of BH9 barrier heights and reaction energies on the percentage of HF exchange for PBEx, $r^2$SCANx, and TPSSx series.

Opposite of what we found for BH9 barrier heights, for reaction energies, adding a dispersion correction significantly improves the accuracy of revPBE and PWPB95 compared to their dispersion uncorrected forms (see Table S7 in the Supporting Information). Our revDSD- and revDOD-PBEP86-D3BJ marginally outperform PWPB95-D3BJ.

Now, what if we gradually increase the percentage of short-range HF exchange while keeping the range separation parameter (ω) fixed? Except for hydride transfer, proton transfer, and nucleophilic substitution reactions, ωDSD functionals prefer 50% or less short-range HF exchange for all other reactions. We have also checked the performance of $\omega DSD_{69}$-PBEP86-D3BJ at five different values of ω ranging from 0.1 to 0.3. Interestingly, for the hydride transfer, proton transfer, and nucleophilic substitution reactions, the range separation parameter has little to no influence on the mean absolute deviations.

**Table 3:** Mean absolute deviations (MAD, in kcal/mol) of different pure and hybrid GGA and meta-GGA functionals for full BH9 barrier height set and its nine subsets. The nine subsets of BH9 are radical rearrangement (I), Diels-Alder (II), halogen atom transfer (III), hydrogen atom transfer (IV), hydride transfer (V), Boron and Silicon-containing reactions (VI), proton transfer (VII), nucleophilic substitution (VIII), and nucleophilic addition (IX).

| Functionals | MAD (kcal/mol) | | | | | | | | | |
|---|---|---|---|---|---|---|---|---|---|---|
| | I | II | III | IV | V | VI | VII | VIII | IX | Total |
| PBE[93,94] | 4.52 | 8.00 | 9.80 | 8.50 | 9.87 | 3.80 | 5.48 | 4.35 | 5.17 | 7.41 |
| PBE10 | 3.08 | 5.69 | 6.69 | 6.24 | 6.99 | 2.85 | 4.36 | 2.26 | 3.80 | 5.26 |
| PBE20 | 1.92 | 3.64 | 4.03 | 4.21 | 4.07 | 2.11 | 3.26 | 1.63 | 2.43 | 3.38 |
| PBE0[110,111] | 1.56 | 2.85 | 2.89 | 3.35 | 2.99 | 1.86 | 2.72 | 1.98 | 1.81 | 2.66 |
| PBE32 | 1.39 | 2.30 | 1.95 | 2.51 | 2.82 | 1.73 | 2.01 | 3.23 | 1.27 | 2.18 |
| PBE38[90] | 1.52 | 2.74 | 2.23 | 2.19 | 3.92 | 1.95 | 1.46 | 4.27 | 1.23 | 2.44 |
| PBE50 | 2.24 | 5.39 | 5.12 | 2.88 | 7.69 | 2.95 | 1.13 | 6.54 | 2.28 | 4.31 |
| PBE60 | 2.87 | 7.89 | 7.69 | 4.30 | 10.92 | 3.78 | 1.41 | 8.18 | 3.51 | 6.19 |
| $r^2$SCAN[112,113] | 3.32 | 7.14 | 8.37 | 7.50 | 10.68 | 2.80 | 3.87 | 4.64 | 4.53 | 6.61 |
| $r^2$SCANh[114] | 2.36 | 4.93 | 5.67 | 5.38 | 7.46 | 1.97 | 2.96 | 2.57 | 3.32 | 4.61 |
| $r^2$SCAN0[114] | 1.35 | 2.32 | 2.20 | 2.59 | 2.79 | 1.19 | 1.77 | 1.19 | 1.61 | 2.12 |
| $r^2$SCAN30 | 1.26 | 1.92 | 1.55 | 2.01 | 2.01 | 1.16 | 1.53 | 1.79 | 1.34 | 1.73 |
| $r^2$SCAN32 | 1.25 | 1.91 | 1.48 | 1.87 | 2.04 | 1.21 | 1.48 | 2.09 | 1.29 | 1.71 |
| $r^2$SCAN35 | 1.27 | 2.05 | 1.68 | 1.74 | 2.43 | 1.34 | 1.41 | 2.58 | 1.23 | 1.81 |
| $r^2$SCAN50[114] | 1.80 | 4.33 | 4.37 | 2.63 | 6.34 | 2.26 | 1.14 | 5.07 | 1.95 | 3.57 |
| $r^2$SCAN60 | 2.26 | 6.53 | 6.45 | 3.94 | 9.43 | 2.86 | 1.38 | 6.62 | 2.90 | 5.21 |
| TPSS[115] | 4.44 | 8.83 | 7.81 | 5.55 | 6.08 | 4.55 | 3.81 | 4.31 | 4.71 | 6.51 |
| TPSSh[116] | 3.16 | 6.59 | 5.01 | 4.01 | 3.86 | 3.83 | 2.81 | 2.46 | 3.39 | 4.68 |
| TPSS0[117] | 1.62 | 3.67 | 2.32 | 2.70 | 3.65 | 2.91 | 1.46 | 2.33 | 1.67 | 2.86 |
| TPSS30 | 1.30 | 3.05 | 2.23 | 2.63 | 4.67 | 2.75 | 1.14 | 2.93 | 1.25 | 2.68 |
| TPSS38 | 1.26 | 3.04 | 3.33 | 2.97 | 6.60 | 2.77 | 1.02 | 4.19 | 1.20 | 3.06 |
| TPSS45 | 1.58 | 4.37 | 4.97 | 3.71 | 8.72 | 3.26 | 1.30 | 5.52 | 1.84 | 4.14 |
| TPSS50 | 1.91 | 5.59 | 6.22 | 4.30 | 10.14 | 3.60 | 1.65 | 6.42 | 2.45 | 5.03 |
| TPSS60 | 2.59 | 8.11 | 8.58 | 5.52 | 12.97 | 4.27 | 2.40 | 8.12 | 3.68 | 6.82 |
| PBE-D3BJ[87,93,94] | 5.22 | 10.01 | 12.74 | 12.02 | 16.20 | 6.42 | 6.11 | 7.80 | 6.36 | 10.09 |
| PBE0-D3BJ[87,110,111] | 2.10 | 3.73 | 4.97 | 6.21 | 7.88 | 3.27 | 3.10 | 1.75 | 2.87 | 4.40 |
| PBE38-D3BJ[87,90] | 1.82 | 2.51 | 1.54 | 3.50 | 3.44 | 2.20 | 1.76 | 1.28 | 2.10 | 2.52 |
| $r^2$SCAN-D3BJ[114] | 3.59 | 8.05 | 9.85 | 9.17 | 14.09 | 4.32 | 4.31 | 6.34 | 5.04 | 7.93 |
| $r^2$SCANh-D3BJ[114] | 2.65 | 5.85 | 7.22 | 7.11 | 11.04 | 3.41 | 3.38 | 4.22 | 3.85 | 5.96 |
| $r^2$SCAN0-D3BJ[114] | 1.62 | 3.04 | 3.38 | 4.23 | 6.28 | 2.38 | 2.13 | 1.40 | 2.20 | 3.29 |
| $r^2$SCAN50-D3BJ[114] | 1.91 | 3.61 | 2.83 | 1.64 | 2.81 | 2.44 | 1.38 | 3.06 | 2.05 | 2.64 |
| TPSS-D3BJ[90,115] | 5.26 | 10.59 | 11.50 | 9.54 | 13.30 | 6.14 | 4.13 | 8.74 | 6.08 | 9.33 |
| TPSSh-D3BJ[90,116] | 3.90 | 7.82 | 8.48 | 7.44 | 10.34 | 4.93 | 3.05 | 6.21 | 4.65 | 7.06 |
| TPSS0-D3BJ[90,117] | 2.30 | 4.14 | 4.05 | 4.49 | 5.41 | 3.22 | 1.60 | 2.77 | 2.60 | 3.86 |

Next, to answer the second research question, we have considered pure and hybrid PBE,[93,94] TPSS,[115] and $r^2$SCAN,[112,113] varying the percentage of exact exchange from 10 to 60% for hybrid GGA and meta-GGA functionals. Considering the 898 barrier heights of BH9, we obtained the best performance near 33% (~1/3) HF exchange for the PBEx and $r^2$SCANx series (where x represents the percentage of exact exchange used in the hybrid functional). However, among the TPSS-based hybrids, TPSS30 offers marginally lower MAD than TPSS33 (see Figure 1, left). For BH9 reaction energies, we get the best performance near 38%, 45%, and 30% HF exchange for the PBE, TPSS, and $r^2$SCAN-based functionals, respectively (see Figure 1, right).

Irrespective of GGA or mGGA functional choice, using a small fraction of HF exchange underestimates and a higher fraction overestimates the barrier heights. On the other hand, the trend is the opposite for the reaction energies (see Table S4 and S5 in the Supporting Information).

For BH9 barrier heights, adding a D3BJ dispersion correction does more harm than good for pure and hybrid functionals with 25% or less HF exchange. This behavior again hints at problems with the suitability of the functional itself. However, for the reaction energies, dispersion corrected forms of these functionals offer better accuracy than the uncorrected ones (see Tables 3 and 4).

**Table 4:** Mean absolute deviations (MAD, in kcal/mol) of different pure and hybrid GGA and meta-GGA functionals for full BH9 reaction energy set and its nine subsets. The nine subsets of BH9 are radical rearrangement (I), Diels-Alder (II), halogen atom transfer (III), hydrogen atom transfer (IV), hydride transfer (V), Boron and Silicon-containing reactions (VI), proton transfer (VII), nucleophilic substitution (VIII), and nucleophilic addition (IX).

| Functionals | MAD (kcal/mol) | | | | | | | | | |
|---|---|---|---|---|---|---|---|---|---|---|
| | I | II | III | IV | V | VI | VII | VIII | IX | Total |
| PBE[93,94] | 2.68 | 8.89 | 6.03 | 3.02 | 2.64 | 4.41 | 2.85 | 2.57 | 2.99 | 5.15 |
| PBE10 | 2.15 | 6.67 | 4.84 | 2.51 | 1.76 | 3.72 | 2.33 | 2.09 | 2.39 | 3.99 |
| PBE20 | 2.05 | 4.76 | 3.85 | 2.10 | 1.49 | 3.02 | 1.84 | 1.61 | 1.91 | 3.07 |
| PBE0[110,111] | 2.16 | 4.04 | 3.52 | 1.95 | 1.47 | 2.70 | 1.63 | 1.44 | 1.75 | 2.75 |
| PBE32 | 2.47 | 3.35 | 3.07 | 1.73 | 1.53 | 2.25 | 1.31 | 1.23 | 1.74 | 2.44 |
| PBE38[90] | 2.80 | 3.33 | 2.79 | 1.60 | 1.75 | 2.02 | 1.10 | 1.22 | 1.86 | 2.42 |
| PBE50 | 3.86 | 4.50 | 2.34 | 1.47 | 2.37 | 1.68 | 0.77 | 1.43 | 2.54 | 2.90 |
| PBE60 | 4.82 | 6.18 | 2.18 | 1.57 | 2.87 | 2.05 | 0.94 | 1.84 | 3.34 | 3.67 |
| $r^2$SCAN[112,113] | 1.39 | 5.64 | 4.36 | 2.07 | 1.71 | 1.67 | 2.21 | 1.69 | 2.58 | 3.28 |
| $r^2$SCANh[114] | 1.33 | 4.21 | 3.57 | 1.75 | 1.59 | 1.47 | 1.81 | 1.50 | 2.23 | 2.63 |
| $r^2$SCAN0[114] | 1.77 | 2.85 | 2.80 | 1.42 | 1.88 | 1.46 | 1.25 | 1.35 | 2.03 | 2.11 |
| $r^2$SCAN30 | 2.02 | 2.64 | 2.63 | 1.35 | 2.06 | 1.54 | 1.10 | 1.32 | 2.03 | 2.06 |
| $r^2$SCAN32 | 2.13 | 2.64 | 2.57 | 1.33 | 2.14 | 1.57 | 1.04 | 1.32 | 2.05 | 2.07 |
| $r^2$SCAN35 | 2.30 | 2.73 | 2.49 | 1.32 | 2.26 | 1.64 | 0.95 | 1.35 | 2.09 | 2.12 |
| $r^2$SCAN50[114] | 3.29 | 4.02 | 2.16 | 1.47 | 2.85 | 2.23 | 0.81 | 1.56 | 2.52 | 2.76 |
| $r^2$SCAN60 | 3.98 | 5.14 | 2.15 | 1.73 | 3.22 | 2.70 | 0.89 | 1.89 | 3.06 | 3.35 |
| TPSS[115] | 3.45 | 12.19 | 5.66 | 2.97 | 2.66 | 6.81 | 2.32 | 2.42 | 4.41 | 6.47 |
| TPSSh[116] | 2.68 | 9.68 | 4.61 | 2.58 | 1.82 | 5.92 | 1.88 | 2.08 | 3.64 | 5.22 |
| TPSS0[117] | 1.95 | 6.16 | 3.47 | 2.21 | 1.31 | 4.54 | 1.28 | 1.77 | 2.34 | 3.60 |
| TPSS30 | 1.97 | 5.10 | 3.17 | 2.14 | 1.31 | 4.06 | 1.08 | 1.73 | 1.92 | 3.17 |
| TPSS38 | 2.21 | 3.91 | 2.83 | 2.06 | 1.51 | 3.40 | 0.90 | 1.69 | 1.42 | 2.71 |
| TPSS45 | 2.63 | 3.54 | 2.55 | 2.03 | 1.84 | 2.83 | 0.88 | 1.71 | 1.15 | 2.57 |
| TPSS50 | 3.02 | 3.67 | 2.39 | 2.03 | 2.10 | 2.53 | 0.94 | 1.79 | 1.19 | 2.65 |
| TPSS60 | 3.99 | 4.78 | 2.33 | 2.13 | 2.67 | 2.29 | 1.08 | 2.08 | 1.97 | 3.20 |
| PBE-D3BJ[87,93,94] | 1.89 | 5.35 | 5.63 | 2.91 | 2.81 | 1.39 | 2.87 | 2.74 | 3.61 | 3.73 |
| PBE0-D3BJ[87,110,111] | 2.43 | 2.35 | 3.10 | 1.84 | 1.25 | 2.20 | 1.65 | 1.46 | 3.44 | 2.23 |
| PBE38-D3BJ[87,90] | 3.52 | 4.12 | 2.40 | 1.54 | 1.56 | 3.03 | 1.10 | 1.45 | 3.98 | 2.88 |
| $r^2$SCAN-D3BJ[114] | 1.27 | 4.38 | 4.23 | 2.03 | 1.71 | 2.01 | 2.23 | 1.81 | 3.16 | 2.92 |
| $r^2$SCANh-D3BJ[114] | 1.33 | 3.02 | 3.42 | 1.71 | 1.51 | 2.21 | 1.83 | 1.65 | 2.95 | 2.33 |
| $r^2$SCAN0-D3BJ[114] | 2.04 | 2.26 | 2.64 | 1.39 | 1.79 | 2.78 | 1.27 | 1.48 | 2.88 | 2.08 |
| $r^2$SCAN50-D3BJ[114] | 3.72 | 5.08 | 1.98 | 1.47 | 2.70 | 4.00 | 0.82 | 1.60 | 3.82 | 3.32 |
| TPSS-D3BJ[90,115] | 2.19 | 7.44 | 5.11 | 2.87 | 2.90 | 2.14 | 2.34 | 2.11 | 2.69 | 4.34 |
| TPSSh-D3BJ[90,116] | 1.60 | 5.17 | 4.08 | 2.50 | 1.98 | 2.03 | 1.90 | 1.85 | 2.26 | 3.26 |
| TPSS0-D3BJ[90,117] | 1.78 | 2.02 | 2.90 | 2.15 | 1.17 | 2.31 | 1.29 | 1.61 | 2.24 | 2.03 |

Closer scrutiny of the performance of nine different reaction energy subsets reveals that for the Diels-Alder reactions, PBE, TPSS, and r$^2$SCAN-based hybrids offer their best performance near 38%, 45%, and 33% HF exchange, respectively. However, for the nucleophilic substitutions, both PBEx and TPSSx series have minima near the same percentage (38%), whereas the r$^2$SCANx series offers the lowest MAD near 33%. A comparatively lower percentage (~20-25%) of HF exchange is preferred by PBEx and r$^2$SCANx series for the radical rearrangements and hydride transfer reactions. For the halogen atom transfer, B & Si-containing reactions, hydrogen atom, and proton transfer reactions, the PBE and TPSS-based hybrids with a relatively large percentage of exact exchanges offer the best accuracy. However, among the r$^2$SCAN-based hybrids, the first two types of reactions prefer a fairly large, and the last two a fairly small percentage, of HF exchange. Finally, the best performers for the nucleophilic addition reactions are PBE32, r$^2$SCAN30, and TPSS45 (see Table 4).

**Table 5:** Mean absolute deviations (MAD, in kcal/mol) of pure and hybrid self-consistent and HF-DFT functionals for BH9 barrier heights and reaction energies. The nine subsets of BH9 are radical rearrangement (I), Diels-Alder (II), halogen atom transfer (III), hydrogen atom transfer (IV), hydride transfer (V), Boron and Silicon-containing reactions (VI), proton transfer (VII), nucleophilic substitution (VIII), and nucleophilic addition (IX).

| Functionals | MAD (kcal/mol) | | | | | | | | | |
|---|---|---|---|---|---|---|---|---|---|---|
| | Barrier Heights | | | | | | | | | |
| | I | II | III | IV | V | VI | VII | VIII | IX | Total |
| PBE[93,94] | 4.52 | 8.00 | 9.80 | 8.50 | 9.87 | 3.80 | 5.48 | 4.35 | 5.17 | 7.41 |
| HF-PBE[118] | 4.43 | 7.93 | 8.25 | 6.69 | 7.62 | 4.75 | 3.42 | 5.53 | 3.52 | 6.62 |
| PBE0[110,111] | 1.56 | 2.85 | 2.89 | 3.35 | 2.99 | 1.86 | 2.72 | 1.98 | 1.81 | 2.66 |
| HF-PBE0[119] | 4.44 | 3.09 | 9.64 | 6.27 | 2.98 | 4.26 | 1.59 | 6.85 | 1.35 | 4.57 |
| PBE-D4[93,94,120] | 5.29 | 10.17 | 12.95 | 12.18 | 16.20 | 6.64 | 6.24 | 8.03 | 6.50 | 10.23 |
| HF-PBE-D4[119] | 3.57 | 9.92 | 6.02 | 5.80 | 14.54 | 3.28 | 3.91 | 2.13 | 4.69 | 7.26 |
| PBE0[110,111,120] | 2.16 | 3.84 | 5.05 | 6.27 | 7.75 | 3.33 | 3.20 | 1.87 | 2.96 | 4.46 |
| HF-PBE0-D4[119] | 3.88 | 4.02 | 6.60 | 4.00 | 6.14 | 2.07 | 1.93 | 2.88 | 2.14 | 4.10 |
| **Functionals** | Reaction Energies | | | | | | | | | |
| PBE[93,94] | 2.68 | 8.89 | 6.03 | 3.02 | 2.64 | 4.41 | 2.85 | 2.57 | 2.99 | 5.15 |
| HF-PBE[118] | 3.77 | 8.46 | 7.14 | 6.90 | 1.37 | 8.16 | 1.75 | 2.49 | 4.21 | 6.24 |
| PBE0[110,111] | 2.16 | 4.04 | 3.52 | 1.95 | 1.47 | 2.70 | 1.63 | 1.44 | 1.75 | 2.75 |
| HF-PBE0[119] | 4.05 | 4.08 | 5.11 | 4.74 | 1.26 | 4.79 | 1.24 | 1.55 | 1.70 | 3.82 |
| PBE-D4[93,94,120] | 1.86 | 4.93 | 5.65 | 2.92 | 2.63 | 1.35 | 2.81 | 2.84 | 3.71 | 3.59 |
| HF-PBE-D4[119] | 3.72 | 3.40 | 6.99 | 6.90 | 1.26 | 3.25 | 1.69 | 2.05 | 2.06 | 4.11 |
| PBE0[110,111,120] | 2.59 | 2.49 | 3.14 | 1.85 | 1.33 | 2.13 | 1.60 | 1.50 | 3.61 | 2.31 |
| HF-PBE0-D4[119] | 4.75 | 3.19 | 4.97 | 4.73 | 1.16 | 2.84 | 1.20 | 1.19 | 2.78 | 3.48 |

Thus far, we have considered the older D3BJ and nonlocal VV10 dispersion corrections; what happens if we use D4 instead, which includes both partial charge dependence and three-body corrections? A small test using ten selected functionals suggests that for BH9 barrier heights, range-separated hybrid and double hybrids do not benefit from substituting D4 for D3BJ (see Table S6 in the Supporting Information). However, global double hybrid functionals, xDSD-PBEP86-D4 and xDOD-PBEP86-D4 perform marginally better than their D3BJ corrected counterparts. For barrier heights, D4 correction does more harm than good for PBE, but B97-D4 performs significantly better than B97-D3BJ. Now, for the BH9 reaction energies, ωB97X-D3BJ performs better than ωB97X-D4. However, using D4 dispersion correction instead of D3BJ has no additional benefit for our DSD-family range separated and global double hybrids. For reaction energies, B97-D3BJ offers better accuracy than B97-D4 (see Table S7 in the Supporting Information).

In previous studies, Sim and Burke,[121,122] the Goerigk group,[84] and the present authors,[119,123] have shown that the use of HF densities instead of self-consistent KS densities can significantly improve the performance of pure and hybrid (with 25% or less exact exchange) GGA and mGGA functionals for noncovalent interactions and barrier heights. Except for the B and Si-containing reactions and nucleophilic substitutions, HF-PBE outperforms its self-consistent counterpart, PBE, for the remaining seven subsets of BH9 barrier heights. However, with D4 dispersion correction HF-PBE-D4 is better than PBE-D4 throughout. Using the HF density does more harm than good for PBE0, but with D4 dispersion correction, HF-PBE0-D4 marginally outperforms PBE0-D4 (see Table 5).

Now, for the BH9 reaction energies, self-consistent functionals perform better than the density-corrected counterparts except for hydride and proton transfer reactions. Using the D4 dispersion correction only reduces the mean absolute error of each functional without affecting the trend (see Table 5).

## IV. Conclusions.

From an extensive survey of global and range-separated hybrid and double hybrid functionals using a large and, more importantly, diverse dataset for barrier heights and reaction energies, we can conclude the following:

- Both for the BH9 barrier heights and reaction energies, B97, $PBE_x$-PBEc, and B88-LYP-family range-separated hybrids functionals outperform their global hybrid counterparts.
- Except for the $PBE_x$-$PBE_c$-family functionals, the range-separated double hybrid functionals perform significantly better than the corresponding global double hybrids for BHs and REs.
- RSX-PBE-QIDH and RSX-PBE-0DH offer better accuracy than the respective global counterparts only for the barrier heights of hydrogen atom transfer reactions. However, among the nine subsets of the BH9 reaction energies, halogen atom transfer, hydrogen atom transfer, and proton transfer reactions are the only three subsets that benefit from range separation in the same family.
- Among all the functionals tested here, the $\omega DOD_{40}$-PBEP86-D3BJ (ω=0.3) and $\omega DOD_{60}$-PBEP86-D3BJ (ω=0.22) are the two best picks (MAD=1.23 kcal/mol) for barrier heights and $\omega DOD_{69}$-PBEP86-D3BJ (ω=0.10) is the best pick for reaction energies overall. Using the more modern D4 instead of D3BJ dispersion correction has no additional benefit. In previous work[36] for the GMTKN55 benchmark, we found that our six-parameter empirical range-separated double hybrids slightly outperform Mardirossian and Head-Gordon's 16-parameter range-separated double hybrid ωB97M(2); for the BH9 set considered here, we find a somewhat more pronounced advantage.

- PBE and $r^2$SCAN-based hybrid functionals offer the lowest mean absolute deviation for BH9 barrier heights near 33% (~1/3) HF exchange, whereas for the TPSSx series it is near 30%. However, for the reaction energies, we obtain the best performance near 38%, 45%, and 30% for the PBEx, TPSSx, and $r^2$SCANx series, respectively.

**Acknowledgments:**
GS acknowledges a doctoral fellowship from the Feinberg Graduate School (WIS). The authors would like to thank Dr. Alberto Otero de la Roza (University of Oviedo, Spain) for supplying the corrected BH9 data ahead of publication.

**Funding Sources:**
This research was funded by the Israel Science Foundation (grant 1969/20) and by the Minerva Foundation (grant 20/05).

**Supporting Information:**
The Supporting Information (in PDF format) is available free of charge at https://doi.org/10.1021/xxxxxxx.
Optimized parameters for our ωDSD-PBEP86-D3BJ, ωDOD-PBEP86-D3BJ and ωnoDispSD-PBEP86 functionals with different fraction of HF exchange and range separation parameter (ω); Root-mean-square deviations (RMSD, kcal/mol) for different DFT functionals on full BH9 barrier height set and its nine subsets; Root-mean-square deviations (RMSD, kcal/mol) for different DFT functionals on full BH9 reaction energy set and its nine subsets; Mean signed deviations (MSD, kcal/mol) for different DFT functionals on full BH9 barrier height set and its nine subsets; and Mean signed deviations (MSD, kcal/mol) for different DFT functionals on full BH9 reaction energy set and its nine subsets.(PDF)

## Table of Contents Graphic:

(8.26cm by 5.0cm)

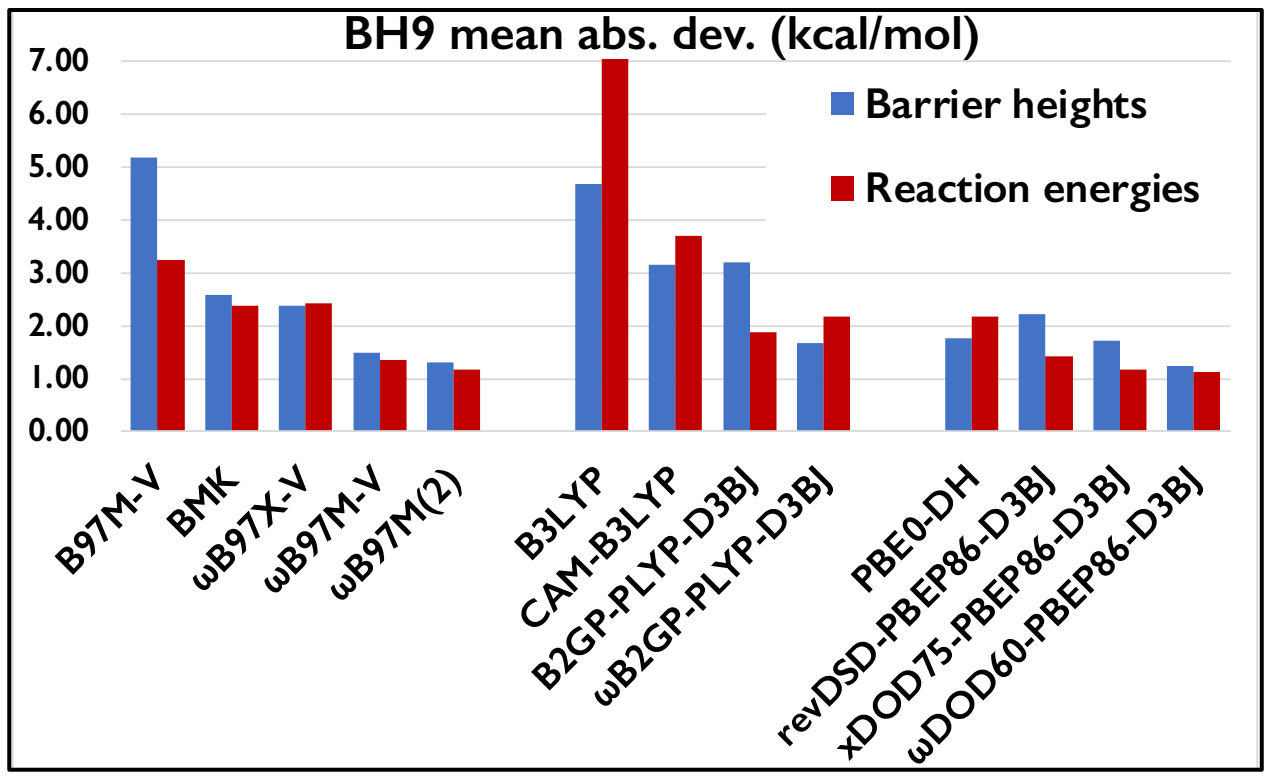